\documentclass[10pt,letterpaper,twocolumn]{article} 

\usepackage{ol2}
\usepackage{hyperref}
\usepackage{amsmath}

\begin{document}

\twocolumn[ 

\title{High performance guided-wave asynchronous heralded single photon source}

\author{O. Alibart, D.B. Ostrowsky and P. Baldi}

\address{Laboratoire de Physique de la Mati\`ere Condens\'ee, UMR
6622 CNRS, Universit\'e de Nice--Sophia Antipolis, Parc Valrose
06108 NICE Cedex 2, France}

\author{S. Tanzilli}
\address{Group of Applied Physics, University of Geneva,\\ 20, rue de
l'Ecole de M\'edecine, 1211 Geneva 4, Switzerland}


\begin{abstract}
We report on a guided wave asynchronous heralded single photon
source based on the creation of non-degenerate photon pairs by
spontaneous parametric down conversion in a Periodically Poled
Lithium Niobate waveguide. We show that using the signal photon at
1310\,$nm$ as a trigger, a gated detection process permits
announcing the arrival of single photons at 1550\,$nm$ at the output
of a single mode optical fiber with a high probability of 0.37. At
the same time the multi-photon emission probability is reduced by a
factor of 10 compared to poissonian light sources. Furthermore, the
model we have developed to calculate those figures of merit is shown
to be very accurate. This study can therefore serve as a paradigm
for the conception of new quantum communication and computation
networks.
\end{abstract}

\ocis{270.5290, 130.2790.}

 ] 

\maketitle

Using the fact that quantum systems are perturbed by measurements
and cannot be cloned with perfect fidelity, Bennett and Brassard
proposed in 1984~\cite{QKD} the possibility of distributing in
absolute confidentiality a cryptographic key between two partners
(commonly called Alice and Bob). Since the security of this protocol
relies on the ability to encode information on only one photon at a
time, single photon sources (SPS) are required. Weak laser pulses
are a very simple solution for approximating SPS behavior but as the
associated photon statistics are poissonian~\cite{mandelwolf} (i.e.
$g^{(2)}(0)$=1), following reference~\cite{secucrypt2}, the
communication is limited to relatively short distances. On the
contrary, an ideal SPS would have a mean number of photons close to
one associated with a $g^{(2)}(0)$=0, allowing a dramatic increase
of the transmission distance.

Quasi-SPS's exhibiting a very low $g^{(2)}(0)$ have already been
demonstrated~\cite{mollpqm,bev2,qdyama02}. While all these solutions
are interesting regarding the photon emission process itself, they
suffer from drawbacks that make them unsuitable for practical
applications at the present time. Furthermore, the emitted photons
of the demonstrated solutions are in the visible spectrum which is
incompatible with communication using installed telecom fiber
networks.

In this letter we report on an alternative solution to the
aforementioned problems. Taking advantage of nonlinear integrated
optics and guided wave technology, we built a heralded single photon
source (HSPS) exhibiting excellent figures of merit and offering a
very practical experimental implementation.

\begin{figure}[htb]
\centerline{\includegraphics[width=8cm]{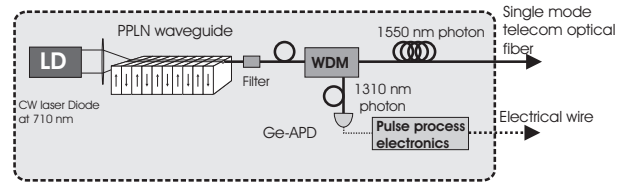}}
\caption{Schematic of our Heralded Single Photon Source. It has an
electrical output which allows heralding the arrival of a single
photon at the single mode optical fiber output.} \label{echt}
\end{figure}

The HSPS relies on photon pairs generated by spontaneous parametric
down-conversion (PDC) in a Periodically Poled Lithium Niobate (PPLN)
optical waveguide. Some of us have previously shown such a structure
to be the most efficient source of down-converted photon pairs
realized to date~\cite{Tanzeleclett}. Since these two photons are
simultaneous to better than 100 fs~\cite{mandelwolf}, the idea is to
use one of them to herald the arrival of the second photon by gating
the associated detector only when it is expected~\cite{pdcmandel}.
This greatly reduces the number of empty pulses that hinder long
distance quantum communication experiments. In our setup the
quasi-phase-matching configuration (for a PPLN period of
13.6\,$\mu\!m$) allows the conversion of pump photons at 710\,$nm$
into pairs of photons whose wavelengths are centered at 1310\,$nm$
and 1550\,$nm$ respectively. As depicted in FIG.~\ref{echt}, taking
advantage of the guided structure, the photon pairs are collected by
a single mode telecom fiber butt-coupled (but not attached) to the
output of the waveguide. After discarding the remaining pump photons
(using a filter in a U-bracket), the pairs are separated by a
standard fiber optic wavelength demultiplexer (WDM). The short
wavelength photons (commonly called ``signal'') are detected using a
$LN_2$ cooled Germanium Avalanche Photodiode (Ge-ADP) operated in
geiger mode with a quantum efficiency of $6\%$. The resulting
electrical signals are used as ``heralds'' for the arrival of the
long wavelength photons (``idler'') which are the expected single
photons. Note that a fiber delay is used to insure the arrival of
these 1550\,$nm$ photons after the heralding electrical pulses at
the two outputs of the HSPS box. Experimentally, the detection of
the single photons amounts to counting a coincidence between the two
photons of the pairs. Therefore, the creation time of the pairs is
not important and this allows pumping the crystal with a CW laser
and having the single photon at 1550\,$nm$ isolated from others
thanks to the gated detection. Essentially, this is a quantum
equivalent of the classical ``asynchronous transfer
mode''~\cite{atm}. Upon receiving the electrical heralding pulses,
the detector is turned on during a given time-window $\Delta T$
(gated mode) ranging from 3\,$ns$ to 50\,$ns$. In this context,
instead of thinking in terms of optical pulses with 0, 1 or several
photons, one has to think in terms of a time-window containing 0, 1
or several photons.

In order to determine the SPS behavior of the 1550\,$nm$ photons, we
wish to measure the probabilities $P_1$ and $P_2$ of having 1 or 2
photons within $\Delta T$ (the probability of having more than 2
photon being negligible). To do this we carried out measurements in
the single photon counting regime using a test bench consisting of a
gated detection Hanbury-Brown $\&$ Twiss type setup~\cite{hanbury}.
In order to evaluate the actual performance of the source, we have
to take into account the poor detection efficiencies, dark counts in
detectors and the fact that $50\%$ of the possible two-photon events
are missed at the beam-splitter which are all parts of the test
bench. Taking these points into account, a correction model proposed
by Mandel~\cite{pdcmandel} has been adapted to calculate $P_1$ and
$P_2$ at the output of the HSPS. The experimental setup was run with
time-windows of 3\,$ns$ and a pump power of 10\,$\mu\!W$. We report
in the first two columns of TABLE~\ref{resultatbrut} the raw data
and the associated experimental probabilities  of finding 1 or 2
photons. Our HSPS exhibits a probability of having an available
single 1550\,$nm$ photon of 0.37 at the output of the single mode
optical fiber. This result is four times better than that of faint
laser pulses and represents to our knowledge one of the best values
for any single photon source. On the other hand the multi-photon
emission probability in a 3\,$ns$ time-window is reduced by a factor
of 10 at equal $P_1$ compared to faint laser pulses. Moreover
TABLE~\ref{resultatbrut} also reports the experimental probabilities
of some existing SPS based on PDC, molecules, NV centers and quantum
dots. Although the comparison is limited to devices for which
explicit values of $P_1$ and $P_2$ have been given, our $P_1$ is
much better than all other techniques when measured at the output of
a collection device such as a lens or an optical fiber and is close
to those using PDC in bulk configuration.

\begin{table*}
\centering \caption{\label{resultatbrut} $P_i$ experimental
probability to find $i=$1 or 2 photons for the HSPS compared to some
of the other existing SPS. Note that, for Quantum Dot, $P_1$
includes the losses at the initial collection lens.}
\begin{tabular}{ll||cccc}
\hline
Data (cts/s) &HSPS &PDC~\cite{fasel}&Molecule~\cite{mollpqm} &NV center~\cite{bev2} &Quantum Dot~\cite{qdyama02}\\
\hline
$N_T=124302$ &$P_1=0.37\pm 0.02$ &$P_1=0.61$ &$P_1=0.047$ &$P_1=0.022$ &$P_1=0.083$\\
Detections$=4702$ &$P_2=0.005\pm 0.001$ &$P_2=2\cdot10^{-4}$ &$P_2=5\cdot10^{-5}$ &$P_2=2\cdot10^{-5}$ &$P_2=4\cdot10^{-4}$\\
 Coincidences$=8$ & $g^{(2)}(0)=0.08\pm 0.02$ &$g^{(2)}(0)=0.002$ &$g^{(2)}(0)=0.046$ &$g^{(2)}(0)=0.07$ &$g^{(2)}(0)=0.14$\\
 \hline
\end{tabular}
\end{table*}

However, our result is quite far from the predicted $100\%$
``preparation efficiency'' announced by Mandel in 1986
\cite{pdcmandel}. This corresponds to the probability to have the
idler photons available when the associated signal photons have been
detected. In practice when a SPS is approximated using PDC, the
limiting factors are the dark counts in the trigger detection and
the losses experienced by the heralded photons. As we measure the
losses of the optical fiber components (filter, WDM and the spool of
hundred meters of standard fiber) to be 1.1\,$dB$ at 1550\,$nm$, we
can then infer the ``preparation efficiency'' $\Gamma$ from the
overall collection efficiency $\gamma$. The latter can be easily
estimated by considering the single detections in the previous setup
after subtraction of the empty states due to the dark counts in the
trigger line.

Calling $N_T$ the raw heralding rate from the Ge-APD and
$D_c\approx20k\,Hz$ its dark counts and noticing that it conditions
the detection of the idler photon, we can write~:
\begin{equation}
\mbox{Single detections
rate}=\gamma\times\eta\times\left(N_T-D_c\right)
\end{equation}
where $\eta$=0.10 represents the quantum efficiency of both
InGaAs-APDs. We experimentally found $\gamma$=0.46. Thus the
preparation efficiency $\Gamma$, which can be seen as the
``guide-to-single mode fiber'' coupling efficiency, was found to be
0.59. As previous experiments based on bulk crystals showed photon
collection efficiencies ranging from 0.03 to
0.83~\cite{ribordy,pdcrarity,fransonsps}, our waveguiding structure
does not improve the best results. However, it is important to note
here that only one butt-coupled fiber is necessary to obtain high
collection efficiency thanks to the collinear PPLN guiding structure
thus offering better stability and ease to use. It is then
interesting to analyze the impact of $\gamma$ on $P_1$ and
$g^{(2)}(0)$ in order to estimate the potential for improvement of
the HSPS.

We begin this analysis by calculating the expected experimental
probability, $P_2$, of having another photon in addition to the
heralded one within a time-window $\Delta T$. As the coherence time
of the single photons ($\tau_c<1$\,$ps$) is much less than the
integration time ($\Delta T\approx$ 3\,$ns$), the number of photons
during $\Delta T$ follows a poissonian distribution and the
probability that the interval from one photon to the next is equal
to or greater than $\Delta T$ is given by~\cite{Poisson}~:
\begin{equation}
P_{\bar{n}}(n=0)=e^{-\gamma\mu\Delta T}
\end{equation}
where $\bar{n}$=$\gamma\mu\Delta T$ is the mean number of photons
per time-window and $\mu$ is the mean emission rate. Considering
that $\gamma\Big(\frac{N_T-D_c}{N_T}\Big)$ is the probability of
having a heralded photon inside $\Delta T$, where $D_c$ is the dark
count rate and $N_T$ the raw counting rate in the Ge-APD, the
probability of having one or more additional photons in this $\Delta
T$ is then~:
\begin{equation}
P_2\approx\gamma^2\mu\Delta
T\left(\frac{N_T-D_c}{N_T}\right)\quad\mbox{with}\quad\mu \Delta
T\ll 1
\end{equation}
Furthermore, the probability of having a single photon inside the
opened time-windows depends on having either a heralded photon or an
additional photon that fills an empty state coming from a dark count
in the trigger detection or from a lost heralded photon~:
\begin{equation}
P_1\approx\gamma\Bigg\{\left(\frac{N_T-D_c}{N_T}\right)\Big[1-2\gamma\mu\Delta\!T\Big]+\mu\Delta\!T\Bigg\}
\end{equation}
Then $g^{(2)}(0)$ corresponds to the ratio between the $P_2$ of a
given SPS and the $P_2=P_1^2/2$ of a poissonian light source running
at equivalent $P_1$~:
\begin{equation}
g^{(2)}(0)\approx\frac{2\mu\:\Delta
T\left(\frac{N_T-D_c}{N_T}\right)}{\Bigg(\mu\Delta
T+\Big[1-2\gamma\mu\Delta\!T\Big]\left(\frac{N_T-D_c}{N_T}\right)\Bigg)^2}
\end{equation}
In our experiment the mean emission rate $\mu$ was estimated taking
into account the detection efficiency of the Ge-APD and the losses
in the trigger line. Taking $\mu$=$6.6\cdot10^{6}$$s^{-1}$ and
$\gamma$=0.46 the first two columns in TABLE~\ref{resultatnet}
report the experimental probabilities and their associated
calculated values. Note the good agreement validating the
theoretical analysis and therefore allowing us to estimate the
performance expected for a better collection efficiency using a
``pigtailed'' fiber. In this particular configuration the fiber is
actually attached to the waveguide using refractive index matching
glue in order to eliminate Fresnel reflections and to fully exploit
the waveguiding structure of the source. $\Gamma$ of 0.7 are
standard in this configuration. Furthermore, efforts underway to
develop new PPLN waveguides enabling down-conversion from 532\,$nm$
to 810\,$nm$ and 1550\,$nm$ could take advantage of passively
quenched silicon detectors showing much higher quantum efficiency
(0.6) for the trigger at 810\,$nm$ and considerably lower noise.
\begin{table}
\centering \caption{\label{resultatnet} $P_i$ probability to find
$i=$1 or 2 photons.}
\begin{tabular}{lll}
\hline
Experimental &calculated &predicted \\
\hline
$P_1=0.37$ &$P_1=0.38\pm0.02$ &$P_1=0.54$\\
$P_2=0.005$ &$P_2=0.003\pm0.001$ &$P_2=7\cdot10^{-4}$\\
$g^{(2)}(0)=0.08$ &$g^{(2)}(0)=0.05\pm0.02$ &$g^{(2)}(0)=0.005$\\
\hline
\end{tabular}
\end{table}

The last column of TABLE~\ref{resultatnet} deals with the predicted
probabilities ($P_1$ and $P_2$) when using a silicon APD and a fiber
pigtailed to the waveguide. In this case, the HSPS would exhibit a
$P_1$ of 0.54, close to the best results reported to date and the
multi-photon emission probability reduced by a factor of 200
compared to usual poissonian light sources at equal $P_1$.

In this letter we have investigated the performance attained using
quasi-phase-matched PDC in a PPLN waveguide associated with optical
fiber components to realize a Heralded Single-Photon-Source at
1550\,$nm$. Using a CW laser we observed a $P_1$ of 0.37 of having a
single photon at the output of a telecom single mode optical fiber,
whereas the multi-photon emission probability is reduced by a factor
of 10 compared to weak laser poissonian light sources at equal
$P_1$. We have also described an accurate model that allows
estimating the expected figures of merit of any asynchronous single
photon source. This, together with the high efficiency previously
reported in~\cite{Tanzeleclett}, demonstrates the potential of
waveguide technologies for building efficient, stable, and compact
sources for quantum communication experiments. Furthermore,
integrated optics could also be used to realize complex passive and
active circuits, permitting a simple implementation of experiments
in the fields of quantum communication and computation.

\section*{Acknowledgments}
We thank the French Ministry of Research through the program
\textit{ACI Photonique} and the STIC Department from CNRS through
\textit{SOQUATOS} project for financial support. One of the authors
(O. Alibart) is grateful to the CNRS and the Regional Council PACA
for their support through a BDI grant.


\email{alibart@unice.fr} \homepage{http://www.unice.fr/lpmc}


\begin{thebibliography}{15}
\expandafter\ifx\csname
natexlab\endcsname\relax\def\natexlab#1{#1}\fi
\expandafter\ifx\csname bibnamefont\endcsname\relax
  \def\bibnamefont#1{#1}\fi
\expandafter\ifx\csname bibfnamefont\endcsname\relax
  \def\bibfnamefont#1{#1}\fi
\expandafter\ifx\csname citenamefont\endcsname\relax
  \def\citenamefont#1{#1}\fi
\expandafter\ifx\csname url\endcsname\relax
  \def\url#1{\texttt{#1}}\fi
\expandafter\ifx\csname urlprefix\endcsname\relax\def\urlprefix{URL
}\fi \providecommand{\bibinfo}[2]{#2}
\providecommand{\eprint}[2][]{\url{#2}}


\bibitem[1]{QKD}
\bibinfo{author}{\bibfnamefont{C.}~\bibnamefont{Bennett}} \bibnamefont{and}
  \bibinfo{author}{\bibfnamefont{G.}~\bibnamefont{Brassard}},
  \bibinfo{journal}{IBM Technical Disclosure Bulletin}
  \textbf{\bibinfo{volume}{28}}, \bibinfo{pages}{3121} (\bibinfo{year}{1985}).

\bibitem[2 ]{mandelwolf}
\bibinfo{author}{\bibfnamefont{L.}~\bibnamefont{Mandel}} \bibnamefont{and}
  \bibinfo{author}{\bibfnamefont{E.}~\bibnamefont{Wolf}},
  \emph{\bibinfo{title}{Optical coherence and quantum optics}}
  (\bibinfo{publisher}{Cambridge university press}, \bibinfo{year}{1995}),
  chap. \bibinfo{chapter}{22.4.7}, pp. \bibinfo{pages}{1084--1088}.

\bibitem[3]{secucrypt2}
\bibinfo{author}{\bibfnamefont{N.}~\bibnamefont{Lutkenhaus}},
  \bibinfo{journal}{Phys.\ Rev.\ A} \textbf{\bibinfo{volume}{61}},
  \bibinfo{pages}{052304} (\bibinfo{year}{2000}).

\bibitem[4]{mollpqm}
\bibinfo{author}{\bibfnamefont{F.}~\bibnamefont{Treussart}},
  \bibinfo{author}{\bibfnamefont{R.}~\bibnamefont{All\'eaume}},
  \bibinfo{author}{\bibfnamefont{V.~L.} \bibnamefont{Floc'h}},
  \bibinfo{author}{\bibfnamefont{L.}~\bibnamefont{Xiao}},
  \bibinfo{author}{\bibfnamefont{J.-M.} \bibnamefont{Courty}},
  \bibnamefont{and} \bibinfo{author}{\bibfnamefont{J.-F.} \bibnamefont{Roch}},
  \bibinfo{journal}{Phys.\ Rev.\ Lett.} \textbf{\bibinfo{volume}{89}},
  \bibinfo{pages}{093601} (\bibinfo{year}{2002}).

\bibitem[5]{bev2}
\bibinfo{author}{\bibfnamefont{A.}~\bibnamefont{Beveratos}},
  \bibinfo{author}{\bibfnamefont{R.}~\bibnamefont{Brouri}},
  \bibinfo{author}{\bibfnamefont{T.}~\bibnamefont{Gacoin}},
  \bibinfo{author}{\bibfnamefont{A.}~\bibnamefont{Villing}},
  \bibinfo{author}{\bibfnamefont{J.-P.} \bibnamefont{Poizat}},
  \bibnamefont{and} \bibinfo{author}{\bibfnamefont{P.}~\bibnamefont{Grangier}},
  \bibinfo{journal}{Phys.\ Rev.\ Lett.} \textbf{\bibinfo{volume}{89}},
  \bibinfo{pages}{187901} (\bibinfo{year}{2002}).

\bibitem[6]{qdyama02}
\bibinfo{author}{\bibfnamefont{M.}~\bibnamefont{Pelton}},
  \bibinfo{author}{\bibfnamefont{C.}~\bibnamefont{Santori}},
  \bibinfo{author}{\bibfnamefont{J.}~\bibnamefont{Vuckovic}},
  \bibinfo{author}{\bibfnamefont{B.}~\bibnamefont{Zhang}},
  \bibinfo{author}{\bibfnamefont{G.~S.} \bibnamefont{Solomon}},
  \bibinfo{author}{\bibfnamefont{J.}~\bibnamefont{Plant}}, \bibnamefont{and}
  \bibinfo{author}{\bibfnamefont{Y.}~\bibnamefont{Yamamoto}},
  \bibinfo{journal}{Phys.\ Rev.\ Lett.} \textbf{\bibinfo{volume}{89}},
  \bibinfo{pages}{233602} (\bibinfo{year}{2002}).

\bibitem[7]{Tanzeleclett}
\bibinfo{author}{\bibfnamefont{S.}~\bibnamefont{Tanzilli}},
  \bibinfo{author}{\bibfnamefont{H.~D.} \bibnamefont{Riedmatten}},
  \bibinfo{author}{\bibfnamefont{W.}~\bibnamefont{Tittel}},
  \bibinfo{author}{\bibfnamefont{H.}~\bibnamefont{Zbinden}},
  \bibinfo{author}{\bibfnamefont{P.}~\bibnamefont{Baldi}},
  \bibinfo{author}{\bibfnamefont{M.~D.} \bibnamefont{Micheli}},
  \bibinfo{author}{\bibfnamefont{D.~B.} \bibnamefont{Ostrowsky}},
  \bibnamefont{and} \bibinfo{author}{\bibfnamefont{N.}~\bibnamefont{Gisin}},
  \bibinfo{journal}{Elec.\ Lett.} \textbf{\bibinfo{volume}{37}}
  (\bibinfo{year}{2001}).

\bibitem[8]{pdcmandel}
\bibinfo{author}{\bibfnamefont{C.}~\bibnamefont{Hong}} \bibnamefont{and}
  \bibinfo{author}{\bibfnamefont{L.}~\bibnamefont{Mandel}},
  \bibinfo{journal}{Phys.\ Rev.\ Lett.} \textbf{\bibinfo{volume}{56}},
  \bibinfo{pages}{58} (\bibinfo{year}{1986}).

\bibitem[9]{atm}
\bibinfo{author}{\bibfnamefont{M.~D.} \bibnamefont{Prycker}},
  \emph{\bibinfo{title}{Asynchronous Transfer Mode: Solution for Broadband
  ISDN}} (\bibinfo{publisher}{Ellis Horwood Ltd}, \bibinfo{year}{1991}).

\bibitem[10]{hanbury}
\bibinfo{author}{\bibfnamefont{R.~H.} \bibnamefont{brown}} \bibnamefont{and}
  \bibinfo{author}{\bibfnamefont{R.~Q.} \bibnamefont{Twiss}},
  \bibinfo{journal}{Nature} \textbf{\bibinfo{volume}{178}},
  \bibinfo{pages}{1046} (\bibinfo{year}{1956}).

\bibitem[11]{fasel}
\bibinfo{author}{\bibfnamefont{S.}~\bibnamefont{Fasel}},
  \bibinfo{author}{\bibfnamefont{O.}~\bibnamefont{Alibart}},
  \bibinfo{author}{\bibfnamefont{S.}~\bibnamefont{Tanzilli}},
  \bibinfo{author}{\bibfnamefont{P.}~\bibnamefont{Baldi}},
  \bibinfo{author}{\bibfnamefont{A.}~\bibnamefont{Beveratos}},
  \bibinfo{author}{\bibfnamefont{N.}~\bibnamefont{Gisin}}, \bibnamefont{and}
  \bibinfo{author}{\bibfnamefont{H.}~\bibnamefont{Zbinden}},
  \bibinfo{journal}{New\ J.\ Phys.} \textbf{\bibinfo{volume}{6}},
  \bibinfo{pages}{163} (\bibinfo{year}{2004})

\bibitem[12]{fransonsps}
\bibinfo{author}{\bibfnamefont{T.}~\bibnamefont{Pittman}},
  \bibinfo{author}{\bibfnamefont{B.}~\bibnamefont{Jacobs}}, \bibnamefont{and}
  \bibinfo{author}{\bibfnamefont{J.}~\bibnamefont{Franson}},
  \bibinfo{journal}{arXiv:quant-ph/0408093}  (\bibinfo{year}{2004}).

\bibitem[13]{ribordy}
\bibinfo{author}{\bibfnamefont{G.}~\bibnamefont{Ribordy}},
  \bibinfo{author}{\bibfnamefont{J.}~\bibnamefont{Brendel}},
  \bibinfo{author}{\bibfnamefont{J.-D.} \bibnamefont{Gautier}},
  \bibinfo{author}{\bibfnamefont{N.}~\bibnamefont{Gisin}}, \bibnamefont{and}
  \bibinfo{author}{\bibfnamefont{H.}~\bibnamefont{Zbinden}},
  \bibinfo{journal}{Phys.\ Rev.\ A} \textbf{\bibinfo{volume}{63}},
  \bibinfo{pages}{012309} (\bibinfo{year}{2000}).

\bibitem[14]{pdcrarity}
\bibinfo{author}{\bibfnamefont{J.}~\bibnamefont{Rarity}},
  \bibinfo{author}{\bibfnamefont{P.}~\bibnamefont{Tapster}}, \bibnamefont{and}
  \bibinfo{author}{\bibfnamefont{E.}~\bibnamefont{Jakeman}},
  \bibinfo{journal}{Optics\ Comm.} \textbf{\bibinfo{volume}{62}},
  \bibinfo{pages}{201} (\bibinfo{year}{1987}).

\bibitem[15]{Poisson}
\bibinfo{author}{\bibfnamefont{R.}~\bibnamefont{Feynman}} \bibnamefont{and}
  \bibinfo{author}{\bibfnamefont{A.}~\bibnamefont{Hibbs}},
  \emph{\bibinfo{title}{Quantum Mechanics and Path Integrals}}
  (\bibinfo{publisher}{McGraw-Hill}, \bibinfo{year}{1965}),
  chap.~\bibinfo{chapter}{12}, p. \bibinfo{pages}{322}.

\end{thebibliography}
\end{document}